\documentclass[preprint,5p,times,sort&compress]{elsarticle}
\usepackage{lineno}
\usepackage{siunitx}
\usepackage{float}

\journal{Nuclear Instruments and Methods in Physics Research A}

\begin{document}

\begin{frontmatter}

\title{Simulation Study of an LWFA-based Electron Injector for AWAKE Run 2}

\author[Manchester,CI]{B. Williamson}
\cortext[mail]{Corresponding author}
\ead{barney.williamson@postgrad.manchester.ac.uk}
\author[Manchester,CI]{G. Xia}
\author[CERN]{S. D{\"o}bert}
\author[MPQ]{S. Karsch}
\author[CERN,MPI]{P. Muggli}

\address[Manchester]{The University of Manchester, Manchester, UK}
\address[CI]{Cockcroft Institute, Warrington, UK}
\address[CERN]{CERN, Geneva, Switzerland}
\address[MPQ]{Max Planck Institute for Quantum Optics, Garching, Germany}
\address[MPI]{Max Planck Institute for Physics, Munich, Germany}

\begin{abstract}
The AWAKE experiment aims to demonstrate preservation of injected electron beam quality during acceleration in proton-driven plasma waves. The short bunch duration required to correctly load the wakefield is challenging to meet with the current electron injector system, given the space available to the beamline. An LWFA readily provides short-duration electron beams with sufficient charge from a compact design, and provides a scalable option for future electron acceleration experiments at AWAKE. Simulations of a shock-front injected LWFA demonstrate a 43 TW laser system would be sufficient to produce the required charge over a range of energies beyond 100 MeV. LWFA beams typically have high peak current and large divergence on exiting their native plasmas, and optimisation of bunch parameters before injection into the proton-driven wakefields is required. Compact beam transport solutions are discussed.  
\end{abstract}


\end{frontmatter}

\section{Introduction}

The Advanced Proton-driven Plasma Wakefield Experiment (AWAKE) \cite{Gschwendtner2016} is a proof-of-principle accelerator at CERN. A 400 GeV proton beam from the CERN Super Proton Synchrotron (SPS) is used to drive GV/m accelerating fields in a 10 m long laser-ionised Rubidium plasma. To resonantly drive wakefields in the plasma a 6-12 cm-long SPS proton bunch is micro-bunched via seeded self-modulation (SSM) \cite{Kumar2010} by co-propagating the protons with the ionising laser pulse. The experimental program of AWAKE consists of measuring and characterising the wakefield driven by an SSM micro-bunched proton beam, and accelerating electrons externally injected into such a wakefield. AWAKE Run 1, lasting until the end of 2018, was initially dedicated to the study of SSM physics and has shown the modulation of the proton beam to be robust. In the remainder of Run I the acceleration of electrons injected from an adjacent beamline, an S-band RF electron gun and booster \cite{Pepitone2016}, will be demonstrated. AWAKE Run 2 will begin in 2021 and aims to show preservation of electron beam emittance and absolute energy spread throughout the acceleration. To achieve this the proton-driven wakefield must be correctly loaded. Recent work has shown that this requires an electron beam with 100-200 pC of charge and a bunch length of 40-60 \si{\micro\meter} \cite{Olsen2018}. Given the small space available to the electron injector beamline such bunch lengths are challenging to meet with the current injector, requiring an upgrade. One proposed solution is to use a laser-driven plasma wakefield accelerator (LWFA) as the electron injector for Run 2  \cite{Muggli2014}, since they readily produce the required beam current as well as high-energy low emittance electron beams in a compact arrangement. This paper studies an LWFA scheme to produce the parameters needed to correctly beam load the proton-driven wakefield, and discusses initial considerations for using such a device at AWAKE.

\section{LWFA electron injector}

The LWFA \cite{Tajima1979} has been well studied since its first proposal, motivated by the utility of a compact source of highly relativistic electrons. An intense laser pulse excites high amplitude electrostatic waves in plasmas - structures sustaining large electric fields that strongly confine and accelerate electrons. The non-relativistic cold plasma wave-breaking limit, which is the maximum amplitude before plasma electron motion overlaps adjacent wave periods, is approximated by $E_0 = m_e \omega_{pe} c / e$, where $\omega_{pe} = (n_e e^2 / \varepsilon_0 m_e)^{1/2}$ is the characteristic angular frequency of oscillation for plasma electrons with density $n_e$. This corresponds to 96 GV/m for a typical plasma density of $10^{18}$ \si{\per\centi\meter\cubed}. Electron beam injection into a laser-driven plasma wakefield has been shown using a broad range of methods, and performed with a variety of plasma sources \cite{Esarey2009}. Efforts have now moved towards the application of LWFA: for example as a source of bright X-rays for high-resolution 3D radiography \cite{Wenz2015}, or, when staged, as a possible linear collider for high-energy physics \cite{Steinke2016}. The injection mechanism used in an LWFA significantly impacts on the final energy distribution, total charge, and emittance of the accelerated electron bunch. Self-injection is the result of a close approach to the wave-breaking limit; a laser pulse intensity and plasma density are selected such that while some charge is injected the accelerating structure is not destroyed. This can lead to strongly peaked GeV-scale electron spectra using a relatively straightforward plasma source \cite{Wang2013}, however once wave-breaking is initiated control over the injection is lost. The threshold at which wave-breaking begins is sensitive to small non-uniformities in the laser driver that vary between shots, leading to significant fluctuations in the quantity and final energy distribution of injected electrons. More reproducible beams of even higher quality are produced when the injection mechanism is separated from the acceleration process. Optical triggering with dedicated laser pulses enables precise control of the injected charge and final energy, and has been used to produce energy spreads of around 1$\%$ \cite{Faure2006}. 
\par
\begin{figure}[htb]\centering
	\includegraphics[width=85mm]{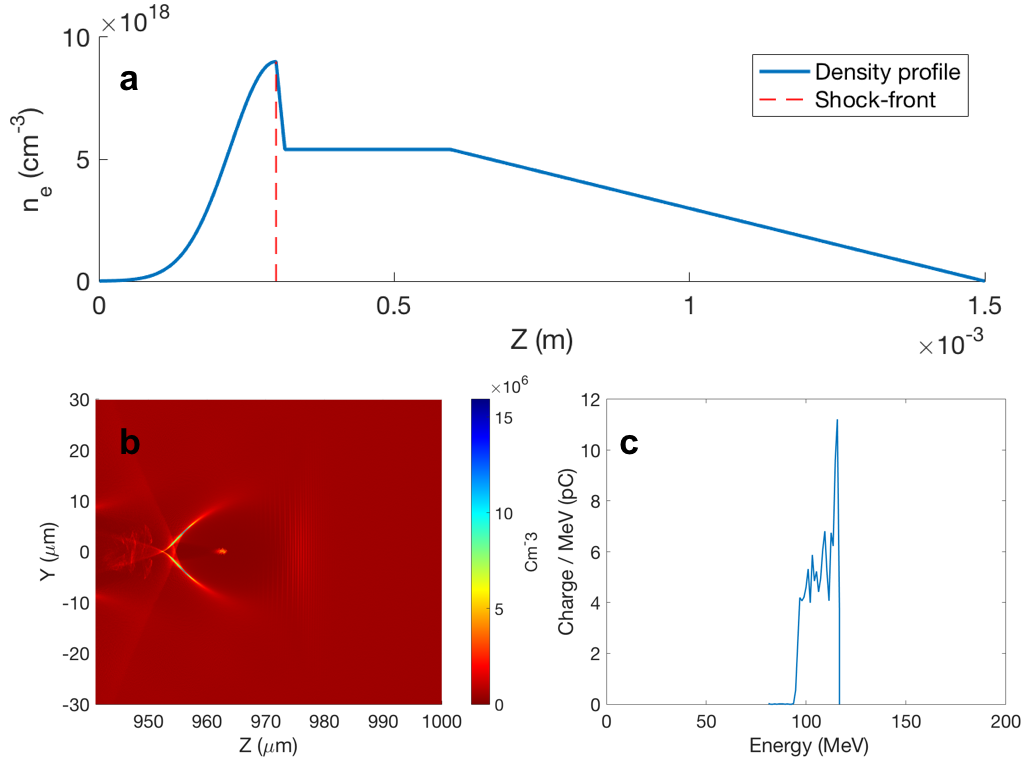}
	\caption{The electron number density $n_e$ used in the simulation to model the supersonic gas jet density profile is shown in \textbf{a}. A charge density plot and energy spectrum of the injected electron bunch at 1 \si{\milli\meter} propagation are shown in \textbf{b} and \textbf{c} respectively to demonstrate the LWFA bubble structure. Z is the direction of laser pulse propagation, perpendicular to the vertical direction Y.}
	\label{fig:composite}
\end{figure}

Shock-front injection \cite{Schmid2010,Buck2013} occurs with a rapid drop in plasma density facilitated by placing an obstacle in the supersonic flow of a Helium gas jet. The resulting hydrodynamic shock forms between two regions of different densities $n_1$ and $n_2<n_1$, where the maximum plasma density $n_1$ is set low enough to inhibit self-injection for a given laser intensity. An idealised longitudinal plasma electron density profile is shown in figure \ref{fig:composite}a. For the presented simulations $n_1$ and $n_2$ correspond to $n_{e,peak}$ and $n_{e,plateau}$ in table \ref{tab:simulationParam}, respectively. If the density transition happens over a distance on the order of a plasma wavelength, $\lambda_{pe} = 2\pi c / \omega_{pe}$, the charge build up behind a propagating laser pulse is rephased into the longer accelerating potential of the new wakefield, formed in the lower density region $n_2$ \cite{Brantov2008}. A fraction of electrons from the first plasma wave behind the laser pulse are effectively injected, typically with a stable absolute energy spread $\Delta E$ and small normalised emittance $\varepsilon_n$ \cite{Buck2013}, because the injection only occurs at the density transition. Within the limit of dephasing, electron energy may be tuned by varying the acceleration length, either by changing the position of the transition in the gas jet, or with gas jets of different sizes. So long as the laser intensity remains low enough to inhibit self-injection for the chosen plasma density, the quantity of injected charge can be varied with the peak value of the laser pulse's normalised vector potential $a_0 \simeq 0.85 \times 10^{-9} \lambda_{\mu m} I^{1/2}$, where $\lambda_{\mu m}$ is the laser wavelength in units of micrometres and $I$ is its intensity in \si{\watt/\centi\metre\squared}. This is because a larger perturbation in plasma electron density follows a stronger laser pulse, so that more electrons are available to be injected at the transition. These characteristics make shock-front injection a scalable source of good quality and reproducible electron beams. 
\begin{table}[htb]\centering
	\centering
	\caption{Basic simulation parameters}
	\begin{tabular}{lcc}
		\hline
		Laser pulse at focus \\
		\hline
		$\lambda$ && 800 \si{\nano\metre} \\
		$a_0$ && 2.3 \\
		$w_0$ (FWHM radius) && 11 \si{\micro\metre} \\
		$\tau$ && 22 fs\\
		\hline
		Plasma \\
		\hline
		$n_{e,peak}$ && $9 \times 10^{18}$  \si{\metre}$^{-3}$ \\ 
		$n_{e,plateau}$ && $5.4 \times 10^{18}$ \si{\metre}$^{-3}$ \\ 
		Total length && 1.5 \si{\milli\metre} \\
		\hline
		Beam parameters after 1.5 mm propagation \\
		\hline
		Charge && 114 pC \\
		$E$ && 112.7 MeV \\
		$\Delta E / E$ && $20.4 \%$ \\\
		$\varepsilon_{n,y}$ && $0.39$ mm mrad \\
		$\sigma_{x,y}$ && 1.5 \si{\micro\meter}  \\
		$\sigma_z$ &&  2.0 \si{\micro\meter}\\
		\hline
	\end{tabular}
	\label{tab:simulationParam}
\end{table}
Particle-in-cell (PIC) simulations in EPOCH \cite{Arber2015a} with two-dimensional slab geometry have been performed to model a shock-front injected LWFA. A range of laser and plasma parameters to produce the required charge for the Run 2 electron injector with a beam energy above 100 MeV are investigated. All simulations have window dimensions of $60 \times 60$ \si{\micro\meter}, 3000 cells in the longitudinal and 600 in the transverse direction respectively, and 4 macroparticles per cell. Modelling an LWFA in such a two-dimensional geometry has two limitations that are relevant to this work. Firstly, the automatic weighting given to each macroparticle is not representative of the total number of electrons present in the laser-plasma interaction: because the electron density is a three-dimensional variable an assumption must be made on the extent of the laser-plasma interaction in the non-simulated third dimension, when estimating the total number of electrons. This is taken to be the FWHM diameter of the laser pulse when estimating the presented bunch charges. Simulations with 2.5 times more macroparticles than the basic simulation show the injected bunch charge to remain at 97\% of the original value. Secondly, relativistic self-focusing of the laser pulse is proportional to $w_0$ in two-dimensions and $w_0^2$ in three. Therefore self-focusing develops at a slower rate in a two-dimensional simulation, and the laser intensity required to produce a given regime of laser-plasma interaction may be overestimated. Table \ref{tab:simulationParam} shows a set of simulation parameters used to produce an electron beam with a sufficient charge of 114 pC for loading the proton-driven wakefield but a bunch length of $\sigma_z = $ 2 \si{\micro\meter} or 6.7 \si{\femto\second}. The resulting peak current of 17 kA will result in beam loading effects on the proton-driven wakefield, causing inefficient acceleration and degradation of witness beam quality. This is a significant challenge of using an LWFA electron injector at AWAKE, and methods to extend $\sigma_z$ must be investigated. 

\begin{figure}[htb]\centering
	\includegraphics[width=90mm]{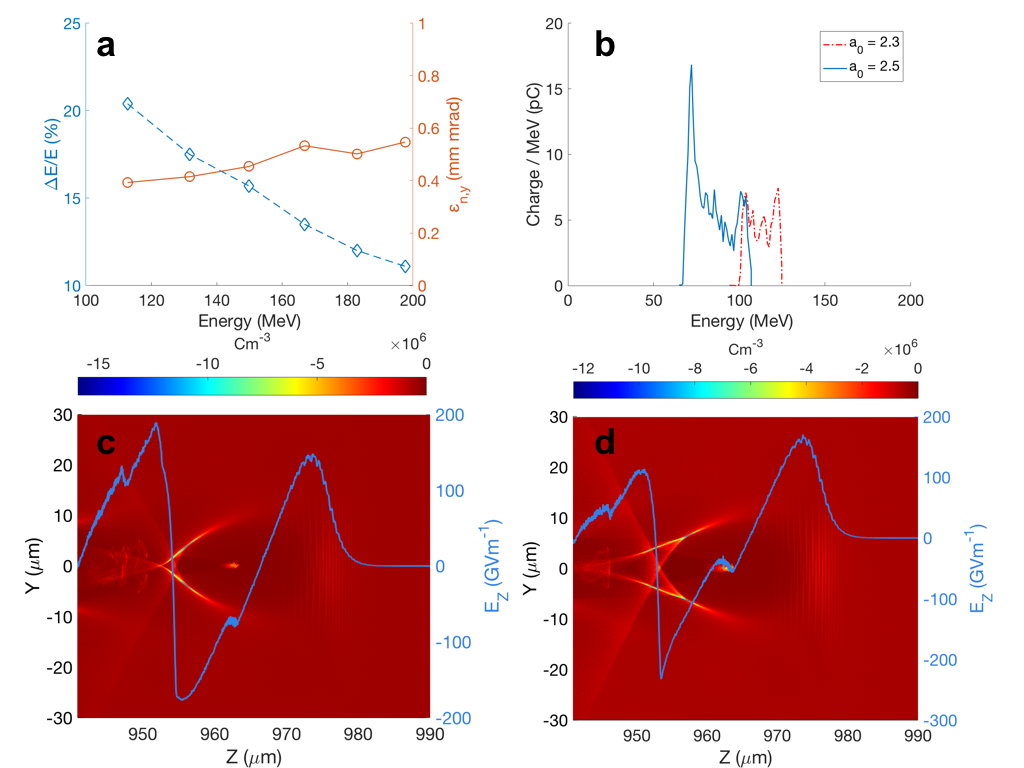}
	\caption{PIC simulation results in \textbf{a} show a falling $\Delta E / E $ ($\diamond$) and small variation in $\varepsilon_{n,y}$($\circ$) for increasing final energy of the injected electron bunch. A broader distribution of energies with a lower average value is seen in \textbf{b} for the injected bunch with $a_0 = 2.5$ (solid line), compared to the basic simulation (dashed line). Longitudinal wakefields are laid over charge density plots for $a_0 = 2.3$ and $a_0 = 2.5$ in \textbf{c} and \textbf{d} respectively. Both indicate beam loading, however in the $a_0 = 2.5$ case $E_z$ shows greater variation along the bunch length, which could result in a larger energy spread within the bunch.}
	\label{fig:composite2}
\end{figure}

An absolute energy spread of $\Delta E \simeq 23$ MeV is found to be stable throughout a range of increasing gas jet size from 1.5 mm to 2 mm in 0.1 mm steps. This results in a relative energy spread $\Delta E / E $ decreasing as $E$ increases, as seen in figure \ref{fig:composite2}a, a trend consistent with experimental observation \cite{Buck2013}. The normalised vertical emittance $\varepsilon_{n,y}$ remains low, below 0.6 mm mrad in the simulated energy range. Excessive beam loading effects on the laser-driven wakefield may contribute to $\Delta E$, as a bunch of charges will modify the wakefield experienced by constituent particles, resulting in a lower overall accelerating gradient that also varies along the bunch. For sufficiently high beam currents such effects become detrimental to acceleration. Figure 2b shows that when $a_0$ is increased from 2.3 to 2.5 a higher total charge of 244 pC is injected, and the bunch now has $E = 83.7$ MeV and $\Delta E / E$ = 46.6\%. The longitudinal electric field $E_z$, shown for both values of $a_0$ in figures \ref{fig:composite2}c,d, is modified by the bunch fields, showing evidence of beam loading. This is more pronounced for $a_0$ = 2.5, due to the higher bunch charge. As a result beam loading effects are likely to contribute to the modified energy spectrum seen in figure 2b and the quality of simulated bunches in general. However, the extent to which beam loading impacts the energy distribution is less clear. A higher laser intensity may be sufficient to initiate wave-breaking and self-injection, a source of additional charge, but interaction between the shock-front and self-injection mechanisms may itself impact on the final energy spectrum when $a_0$ is increased. Nonetheless beam loading may be optimised to reduce $\Delta E$. Owing to the simulation geometry a derived charge density for the beam will not be representative of the real value, making it challenging to evaluate beam loading effects precisely.
\section{Beam transport considerations}

\begin{figure}[htb]\centering
	\includegraphics[width= 85mm]{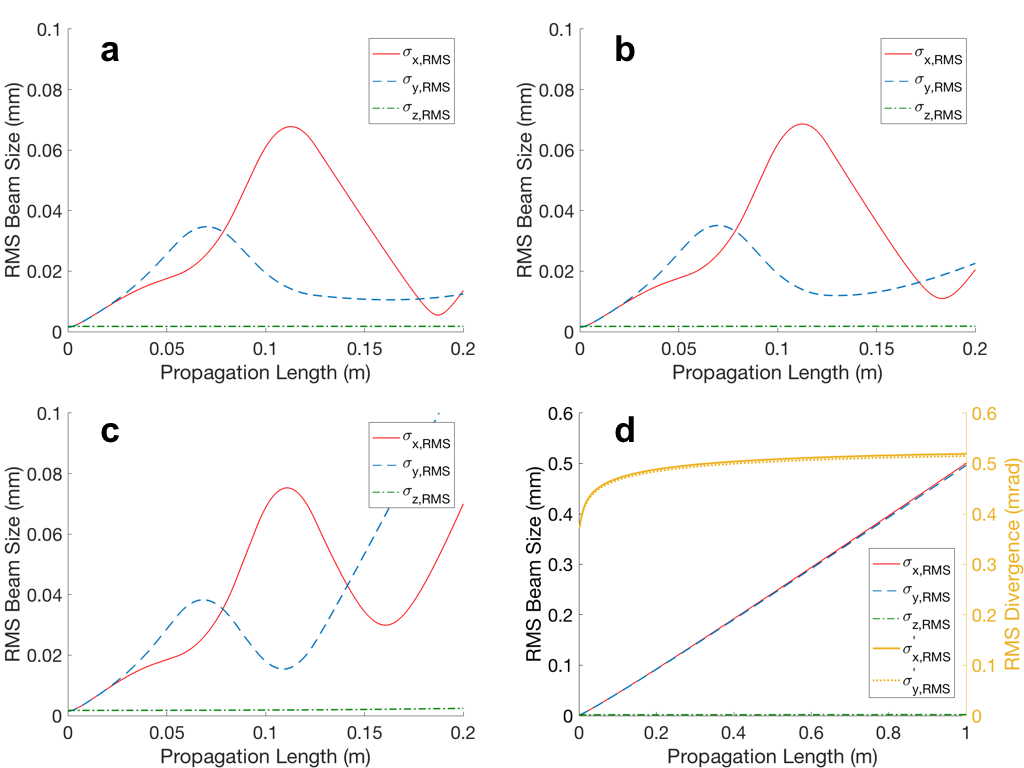}
	\caption{Transverse beam size evolution of an electron bunch from the basic simulation during focusing in a PMQ triplet. PMQs are positioned at 0.03, 0.07, and 0.11 m. Only the absolute energy spread is varied between plots; $\Delta E = 5$ MeV in \textbf{a}, $\Delta E = 10$ MeV in \textbf{b}, $\Delta E = 23$ MeV in \textbf{c}. Free propagation over 1 m in vacuum of a bunch with $\Delta E = 23$ MeV is shown in \textbf{d}.}
	\label{fig:tripletFocusing} 
\end{figure}

It is important to understand how the electron beam parameters evolve once leaving the gas jet so that they can be optimised prior to injection. The laser-driven wakefield provides strong linear transverse focusing of the electron beam. On exiting the plasma this is lost and the beam diverges due to its finite emittance. An electron beam from the basic simulation is implemented in the space-charge tracking code ASTRA \cite{Floettmann2017}. Figure \ref{fig:tripletFocusing}d shows the beam's free propagation in vacuum tracked over a metre. Transverse beam sizes $\sigma_{x,y}$ grow significantly, and while the bunch duration $\sigma_z$ increases slightly it remains close to its value at generation. Tracking was performed both with and without space-charge forces, demonstrating $\varepsilon_{n,y}$ growth in both cases. Partly this is accounted for by space-charge effects, although the intrinsic growth of $\varepsilon_n$ during free propagation is seen in electron beams with large energy spread and divergence \cite{Migliorati2013}. As $\sigma_z$ remains small the need for longitudinal phase space manipulation of the electron beam before injection into the proton-driven wakefield is more apparent. Focusing is also needed to reduce the transverse beam sizes, ideally to 5.25 \si{\micro\meter}, used in AWAKE electron injection studies \cite{Olsen2018}. In Run 2 the AWAKE plasma source will be replaced with two separate vapour vessels, one for modulating the proton beam and one for acceleration. This presents an opportunity to place the LWFA injector in the space between, on-axis with the proton beam. Because space is limited, to 1 m or less to avoid proton beam divergence and a sub-optimal accelerating wakefield, injector components should be compact. To facilitate symmetric focusing a quadrupole triplet could be placed after the LWFA. The thin lens approximation is valid if the length $l$ of a quadrupole is small compared to its focal length $f$, and in this case the strength of a quadrupole is given by $k = 1/(fl) = (eg)/p$ where $g$ is the magnetic gradient and $e$ and $p$ are the charge and momentum respectively of a particle within the magnetic field. Alternatively:

\begin{equation}
k [m^{-2}] = 0.2998\frac{g [T/m]}{p [GeV/c]}
\end{equation}

For an electromagnetic quadrupole with $g = 15$ T/m and $l = 20 $ cm the focal length for a 113 MeV electron beam is 12.6 cm, and over a metre in a triplet configuration. Focal lengths could be decreased with larger magnets but again this would exceed the spatial restraints of an on-axis LWFA injector. Additionally space for final focusing optics would be required to align the drive laser using, for example, plasma-mirror tape \cite{Steinke2016}. One possibility to achieve compact focusing is with permanent magnetic quadrupoles (PMQs), since they have magnetic gradients up to 560 T/m \cite{Lim2005} and cm-scale focal lengths. A simple magnetic lattice was added to the simulation in ASTRA consisting of three PMQ lenses of 236 T/m, 539 T/m, and 471 T/m, each 30 mm in length. Focused beam sizes for a range of electron bunch energy spreads are plotted in figures \ref{fig:tripletFocusing}a-c.  

The energy dependence of quadrupole focusing can be seen in equation 1, and this chromatic behaviour results in a broad distribution of foci for large energy spreads. For $\Delta E/E = 5\%$ beam sizes are focused to $\sigma_{x,y}$ = 10.8 \si{\micro\metre} in 20 \si{\centi\metre} whereas for $\Delta E/E = 20.4\%$, found in the basic simulation, focusing is no longer symmetric and beam waists are larger. Individual magnet strength and positioning within the triplet must be optimised to reach $\sigma_{x,y}$ = 5.25 \si{\micro\metre}, but ultimately energy spread should be reduced at generation to aid beam transport. Specifications for a series of dispersive elements to stretch $\sigma_z$ from 2 \si{\micro\meter} to the required range of 40-60 \si{\micro\meter} are currently being studied. The triplet in figure \ref{fig:tripletFocusing} shows a possible minimum focal length that can be increased to accommodate the required dispersive elements. Detailed design for the entire beam transport lattice will be performed with the tracking code Elegant, allowing for coherent synchrotron radiation modelling, relevant for transport of short duration electron beams.
 
\section{Summary}

A set of laser and plasma parameters for a shock-front injected LWFA have been investigated as a possible electron injector for Run 2 at the AWAKE experiment. PIC simulations show charge requirements can be met with a laser of $a_0 = 2.3$, corresponding to a power of 43 TW for $\tau=$ 22 fs and $w_0=$ 11 \si{\micro\meter}. Such an LWFA is scalable in energy and charge and performs with small variations in $\Delta E$ and $\varepsilon_n$, resulting in good quality electron beams with decreasing $\Delta E / E$ for higher energies. Future PIC studies will improve beam quality at generation. Experimentally, doping of the Helium gas target with a small fraction of Nitrogen has been shown to minimise the impact of shot-to-shot variations in laser profile on accelerated electron beam spectra, from shock-front injected LWFA \cite{Thaury2015}. After generation the electron bunch diverges strongly but retains its short duration. A basic PMQ triplet has been investigated as a source of compact focusing prior to injection into the proton-driven wakefield. A detailed design for the entire transport beamline of the LWFA injector, including a dispersive section to lower peak beam current, is currently being produced.     

\section*{Acknowledgements}

This work was supported by the Cockcroft Institute Core Grant, and the Science and Technologies Facilities Council (STFC). Computing resources provided by STFC Scientific Computing Department's SCARF cluster. The authors would like to thank Andreas D{\"o}pp for helpful discussions.

\bibliography{mybibfile}

\end{document}